\def\ph{\phantom{-}}  
\def\mb#1{\mbox{\boldmath{$#1$}}}
\documentclass[amsmath,amssymb,superscriptaddress,showkeys, showpacs, aps, nofootinbib]{revtex4}
\usepackage{graphicx}
\usepackage{units}
\usepackage{amstext}
\usepackage{mathrsfs} 
\usepackage{cases}
\begin{document}

%\hspace*{4.0 in}[drcomsp.tex: 15{$^{\rm th}$} January, 2016]
\hspace*{4.5 in} CUQM - 155
\vspace{0.5 in}
\markboth{R.~L.~Hall \& P. Zorin}{Refined comparison theorems for the Dirac equation with spin and pseudo--spin symmetry in $d$ dimensions.}

%%%%%%%%%%%%%%%%%%%%%%%%%%%%%%%%%%%%%%%%%%%%%%%%%%%%%%%%%%%%%%%%%%%%%%%%%%%%
\title{Refined comparison theorems for the Dirac equation with spin and pseudo--spin symmetry in $d$ dimensions.}
%%%%%%%%%%%%%%%%%%%%%%%%%%%%%%%%%%%%%%%%%%%%%%%%%%%%%%%%%%%%%%%%%%%%%%%%%%%%

\author{Richard L. Hall}
\email{richard.hall@concordia.ca}
\affiliation{Department of Mathematics and Statistics, Concordia University,
1455 de Maisonneuve Boulevard West, Montr\'{e}al,
Qu\'{e}bec, Canada H3G 1M8}

\author{Petr~Zorin}
\email{petrzorin@yahoo.com}
\affiliation{Department of Mathematics and Statistics, Concordia University,
1455 de Maisonneuve Boulevard West, Montr\'{e}al,
Qu\'{e}bec, Canada H3G 1M8}

%%%%%%%%%%%%%%%%%%%%%%%%%%%%%%%%%%%%%%%%%%%%%%%%%%%%%%%%%%%%%%%%%%%%%%%%%%%
%%%%%%%%%%%%%%%%%%%%%%%%%%%%%%%%%%%%%%%%%%%%%%%%%%%%%%%%%%%%%%%%%%%%%%%%%%%
\begin{abstract} 
The classic comparison theorem of quantum mechanics states that if two potentials are ordered then the corresponding energy eigenvalues are similarly ordered, that is to say if $V_a\le V_b$, then $E_a\le E_b$. Such theorems have recently been established for relativistic problems even though the discrete spectra are not easily characterized variationally. In this paper we improve on the basic comparison theorem for the Dirac equation with spin and pseudo--spin symmetry in $d\ge 1$ dimensions. The graphs of two comparison potentials may now cross each other in a prescribed manner implying that the energy values are still ordered. The refined comparison theorems are valid for the ground state in one dimension and for the bottom of an angular momentum subspace in $d>1$ dimensions. For instance in a simplest case in one dimension, the condition $V_a\le V_b$ is replaced by $U_a\le U_b$, where $U_i(x)=\int_0^x V_i(t)dt$, $x\in[0,\ \infty)$, and $i=a$ or $b$. 
\end{abstract}
%%%%%%%%%%%%%%%%%%%%%%%%%%%%%%%%%%%%%%%%%%%%%%%%%%%%%%%%%%%%%%%%%%%%%%%%%%%
%%%%%%%%%%%%%%%%%%%%%%%%%%%%%%%%%%%%%%%%%%%%%%%%%%%%%%%%%%%%%%%%%%%%%%%%%%%

\keywords{Dirac equation, ground state, spin symmetry, pseudo--spin symmetry, comparison theorems, refined comparison theorems.}

\pacs{03.65.Pm, 03.65.Ge, 36.20.Kd.}

\maketitle

%%%%%%%%%%%%%%%%%%%%%%%%%%%%%%%%%%%%%%%%%%%%%%%%%%%%%%%%%%%%%%%%%%%%%%%%%%%%%%%%%%%%%%%%
%%%%%%%%%%%%%%%%%%%%%%%%%%%%%%%%%%%%%%%%%%%%%%%%%%%%%%%%%%%%%%%%%%%%%%%%%%%%%%%%%%%%%%%%
\section{Introduction}
%%%%%%%%%%%%%%%%%%%%%%%%%%%%%%%%%%%%%%%%%%%%%%%%%%%%%%%%%%%%%%%%%%%%%%%%%%%%%%%%%%%%%%%%
%%%%%%%%%%%%%%%%%%%%%%%%%%%%%%%%%%%%%%%%%%%%%%%%%%%%%%%%%%%%%%%%%%%%%%%%%%%%%%%%%%%%%%%%
Spin and pseudo--spin symmetry were first introduced in \cite{SSV1, SSV2} more than forty years ago. Spin symmetry occurs in the spectrum of a meson \cite{Mes, Nu2, Nu3, Nu4}. Pseudo--spin symmetry helps explain the spectra of deformed nuclei \cite{Nuc1} and superdeformation \cite{Sup}, which occurs in the spectra of certain nuclei \cite{Nuc2}. Spin symmetry helps in the design of nuclear shell--model schemes \cite{NS1, NS2, NS3}, and is used to explain certain identical bands \cite{IB1, IB2, IB3}.  Exact spin symmetry in the Dirac equation occurs when the difference between the scalar $S$ and vector $V$ potentials is equal to a constant, i.\,e. $S-V=c_1$ \cite{Nu2}. While exact pseudo--spin symmetry exists when the sum of scalar $S$ and vector $V$ potentials is equal to a constant, i.\,e. $S+V=c_2$ \cite{S-V1, S-V2}. Here we consider potentials of equal magnitude, so that $|S|=|V|$, and the constants $c_1$ and $c_2$ are zero.

Under spin or pseudo--spin symmetries a relativistic system of Dirac coupled equations can be written as a single Schr\"{o}dinger--like equation. Then one can use methods which were developed to solve non--relativistic equations exactly or approximately, such as factorization and path--integral methods \cite{FPM1, FPM2, FPM3, FPM4, FPM5}, the Nikiforov--Uvarov method \cite{NU}, shape invariance \cite{SI1, SI2}, asymptotic iteration method \cite{AIM1, AIM2, AIM3, AIM4, AIM5}, supersymmetric quantum mechanics \cite{SU}, and so on. For instance, the Dirac equation was solved for the Morse potential \cite{Mor1, Mor2, Mor3, Mor4, Mor5}, the harmonic--oscillator potential \cite{Har1, Har2, Har3}, the pseudoharmonic potential \cite{PSH}, the P\"{o}schl--Teller potential \cite{PT1, PT2, PT3, PT4}, the Woods--Saxon potential \cite{WS1, WS2}, the Eckart potential \cite{Eck1, Eck2}, the Coulomb and the Hartmann potentials \cite{CH}, the Hyperbolic potentials and the Coulomb tensor interaction \cite{Hyp1, Hyp2}, the Rosen--Morse potential \cite{RM}, the Hulth\'en potential \cite{Hul1, Hul2, Hul3}, the Hulth\'en potential including the Coulomb--like tensor potential \cite{Hul0}, the $v_0\tanh^2(r/d)$ potential \cite{Pot}, the Coulomb--like tensor potential \cite{CT}, the modified Hylleraas potential \cite{Hyl}, the Manning--Rosen and the generalized Manning--Rosen potentials \cite{MR1, MR2, MR3, MR4, MR5}, and others.  The point is that there are many known exact solutions that can be used for comparisons with new potentials found in given problems. 

The comparison theorem of quantum mechanics states that if the comparison potentials are ordered then the corresponding energy eigenvalues are ordered as well, i.\,e. if $V_a\le V_b$ then $E_a\le E_b$ \cite{CT1, CT2, CT3, CT4, CT5, CT6, CT7}, thus the graphs of the comparison potentials are not allowed to cross over each other. The comparison theorem was also established for the Dirac equation under the spin and pseudo--spin symmetry \cite{HY}. Similarly to the non--relativistic case \cite{Hall7}, here we refine the comparison theorem for the Dirac equation under the spin and pseudo--spin symmetry by establishing conditions under which the potentials can intersect and still preserve the ordering of eigenvalues. In the simplest one--dimensional case, the condition $V_a\le V_b$ is replaced by $U_a\le U_b$, where $U_i(x)=\int_0^x V_i(t)dt$, $x\in[0,\ \infty)$, and $i=a$ or $b$.

The paper is organized in the following way: we start with the Dirac equation in one dimension and derive the usual comparison theorem (section II. A.). Then in section II. B. we establish some general relations between the  potential $V$,  the energy $E$, and mass of the particle $m$. In section II. C. we refine the comparison theorem, using necessary monotone behaviour of the wave functions. Finally, we demonstrate how to apply the refined comparison theorems in practice, often by taking advantage of the corollaries with specially designed simplified sufficient conditions (section II. D.). Following a similar path we then consider the family of $d>1$ dimensional cases. In order to simplify the statements and proofs of the theorems, we shall usually combine the formulation of the spin--symmetric and pseudo--spin-symmetric cases by the use of a parameter $s=\pm 1.$

%%%%%%%%%%%%%%%%%%%%%%%%%%%%%%%%%%%%%%%%%%%%%%%%%%%%%%%%%%%%%%%%%%%%%%%%%%%%%%%%%%%%%%%%
%%%%%%%%%%%%%%%%%%%%%%%%%%%%%%%%%%%%%%%%%%%%%%%%%%%%%%%%%%%%%%%%%%%%%%%%%%%%%%%%%%%%%%%%
\section{The one--dimensional case $d=1$.}
%%%%%%%%%%%%%%%%%%%%%%%%%%%%%%%%%%%%%%%%%%%%%%%%%%%%%%%%%%%%%%%%%%%%%%%%%%%%%%%%%%%%%%%%
%%%%%%%%%%%%%%%%%%%%%%%%%%%%%%%%%%%%%%%%%%%%%%%%%%%%%%%%%%%%%%%%%%%%%%%%%%%%%%%%%%%%%%%%
%%%%%%%%%%%%%%%%%%%%%%%%%%%%%%%%%%%%%%%%%%%%%%%%%%%%%%%%%%%%%%%%%%%%%%%%%%%%%%%%%%%%%%%%
%%%%%%%%%%%%%%%%%%%%%%%%%%%%%%%%%%%%%%%%%%%%%%%%%%%%%%%%%%%%%%%%%%%%%%%%%%%%%%%%%%%%%%%%
\subsection{The Dirac equation}
%%%%%%%%%%%%%%%%%%%%%%%%%%%%%%%%%%%%%%%%%%%%%%%%%%%%%%%%%%%%%%%%%%%%%%%%%%%%%%%%%%%%%%%%
%%%%%%%%%%%%%%%%%%%%%%%%%%%%%%%%%%%%%%%%%%%%%%%%%%%%%%%%%%%%%%%%%%%%%%%%%%%%%%%%%%%%%%%%
The Dirac equation in one dimension is given by \cite{calog}: 
\begin{equation*}
\left(\sigma_1\frac{\partial}{\partial x}-(E-V)\sigma_3+m+S\right)\psi=0,
\end{equation*}
in natural units $\hbar=c=1$, $m$ is the mass of the particle, and $\sigma_1$ and $\sigma_3$ are Pauli matrices. The potentials $V$ and $S$ are monotone even functions such that the energy eigenvalue $E$ exists. Both potentials are bounded at the origin, that is to say $V(0)$ and $S(0)$ are finite. By taking the two--component Dirac spinor as $\psi=\left(\begin{array}{cc}\varphi_1 \\ \varphi_2\end{array}\right)$ the above matrix equation can be decomposed into the following system of first--order linear differential equations \cite{Dombey, Qiong}:
\begin{subnumcases}{}
\label{d1l}
\varphi_1'=-(E+m-V+S)\varphi_2,\\
\label{d1r}
\varphi_2'=\ph(E-m-V-S)\varphi_1,
\end{subnumcases}
where the prime $\prime$ denotes the derivative with respect to $x$. For bound states, $\varphi_1$ and $\varphi_2$ satisfy the normalization condition
\begin{equation*}
(\varphi_1,\varphi_1) + (\varphi_2,\varphi_2) = \int\limits_{-\infty}^{\infty}(\varphi_1^2 + \varphi_2^2)dx = 1.
\end{equation*}

We now compare two problems with potentials $V_i$ and $S_i$, $i=a$ or $b$, and corresponding energies $E_a$ and $E_b$ for which the system (\ref{d1l})--(\ref{d1r}) becomes respectively
\begin{subnumcases}{}
\label{3}
\varphi_{1a}'=-(E_a+m-V_a+S_a)\varphi_{2a},\\
\label{4}
\varphi_{2a}'=\ph(E_a-m-V_a-S_a)\varphi_{1a},
\end{subnumcases}
and
\begin{subnumcases}{}
\label{5}
\varphi_{1b}'=-(E_b+m-V_b+S_b)\varphi_{2b},\\
\label{6}
\varphi_{2b}'=\ph(E_b-m-V_b-S_b)\varphi_{1b}.
\end{subnumcases}
Let us consider the following combination of the above equations: 
\begin{equation*}\label{comb}
\text{(\ref{3})}\varphi_{2b} - \text{(\ref{4})}\varphi_{1b} - \text{(\ref{5})}\varphi_{2a} + \text{(\ref{6})}\varphi_{1a},
\end{equation*}
which, after some simplifications, becomes
\begin{equation*}
(\varphi_{1a}\varphi_{2b})'-(\varphi_{2a}\varphi_{1b})'=(\varphi_{1a}\varphi_{1b}+
\varphi_{2a}\varphi_{2b})(E_b-E_a-V_b+V_a)-(\varphi_{1a}\varphi_{1b}-
\varphi_{2a}\varphi_{2b})(S_b-S_a).
\end{equation*}
Integrating the left side of the above expression by parts from $0$ to $\infty$, and using the boundary conditions, we find $\int_0^\infty \left[(\varphi_{1a}\varphi_{2b})'-(\varphi_{2a}\varphi_{1b})'\right]dx=0$. 
We then integrate the right side to obtain
\begin{equation}\label{expr1}
(E_b-E_a)\int_0^\infty (\varphi_{1a}\varphi_{1b}+\varphi_{2a}\varphi_{2b})dx=
\int_0^\infty\left[ (S_b+V_b-S_a-V_a)\varphi_{1a}\varphi_{1b}+
(S_a-V_a-S_b+V_b)\varphi_{2a}\varphi_{2b}\right]dx.
\end{equation}
We can merge the spin and pseudo--spin symmetric cases (as was done in \cite{HY}) by introducing the parameter $s$, which is equal to $1$ if $S=V$ and $-1$ if $S=-V$, so $S=sV$. Then the above expression becomes
\begin{equation}\label{VS}
(E_b-E_a)\int_0^\infty (\varphi_{1a}\varphi_{1b}+\varphi_{2a}\varphi_{2b})dx=
2\int_0^\infty (V_b-V_a)\varphi_{qa}\varphi_{qb}dx,
\end{equation}
where $q=1$ if $s=1$ and $q=2$ if $s=-1$. Expression (\ref{VS}) yields spectral ordering if the comparison potentials are ordered and the integrands have constant signs, i.\,e. $E_a\le E_b$ if $V_a\le V_b$. This is equivalent to the comparison theorem \cite{HY} which was derived by Hall and Ye\c{s}ilta\c{s} using monotonicity properties and is valid also for exited states. However, the potentials are not allowed to crossover. In the present paper we refine this theorem by letting the potentials intersect each other in a suitable controlled manner and still imply spectral ordering.
%%%%%%%%%%%%%%%%%%%%%%%%%%%%%%%%%%%%%%%%%%%%%%%%%%%%%%%%%%%%%%%%%%%%%%%%%%%%%%%%%%%%%%%%
%%%%%%%%%%%%%%%%%%%%%%%%%%%%%%%%%%%%%%%%%%%%%%%%%%%%%%%%%%%%%%%%%%%%%%%%%%%%%%%%%%%%%%%%
\subsection{Classes of potentials}
%%%%%%%%%%%%%%%%%%%%%%%%%%%%%%%%%%%%%%%%%%%%%%%%%%%%%%%%%%%%%%%%%%%%%%%%%%%%%%%%%%%%%%%%
%%%%%%%%%%%%%%%%%%%%%%%%%%%%%%%%%%%%%%%%%%%%%%%%%%%%%%%%%%%%%%%%%%%%%%%%%%%%%%%%%%%%%%%%
By differentiation and substitution, system (\ref{d1l})--(\ref{d1r}) in the case $S=sV$ can be written as a Schr\"{o}dinger--like equation
\begin{equation}\label{Sch}
-\varphi '' + 2V(E+sm)\varphi=(E^2-m^2)\varphi,
\end{equation}
where $\varphi=\varphi_1$ if $s=1$ and $\varphi=\varphi_2$ if $s=-1$. The radial function $\varphi$ is normalizable but not necessarily normalized. In any case, and the above eigenequation determines, the eigenvalue $E$. By using the spectral properties of the Schr\"{o}dinger operator \cite{Reed}, we propose 
to consider two subclasses of potentials: $(i)$ $V$ is finite for large $|x|$ and without loss of generality we choose the energy scale so that $\lim\limits_{|x|\to\infty}V=0$; and $(ii)$ $V$ is unbounded for large $|x|$ and without loss of generality we choose a coordinate system so that $V(0)=0$. Analysing (\ref{Sch}) and (\ref{d1l})--(\ref{d1r}) for the $S=sV$ case we can finally state the three classes of potentials and corresponding relationship between energy $E$ and mass $m$:\\ 

{\it $(i)$ $V$ is finite near infinity, $sV(0)<0$, and\\

\hspace{3cm} (1) $sV\le 0$ and $\lim\limits_{|x|\to\infty}V=0$. This implies $-m<E<m$;\\

$(ii)$ $V$ is unbounded near infinity, $V(0)=0$, and\\

\hspace{3cm} (2) $sV\ge 0$ and $\lim\limits_{|x|\to\infty}V=\ph s\infty$. This implies $sE>\ph m$\\

or\\

\hspace{3cm} (3) $sV\le 0$ and $\lim\limits_{|x|\to\infty}V=-s\infty$. This implies $sE<-m$.}\\

For instance, consider $s=-1$ case. Then it follows from (\ref{Sch}) that if $V\le 0$ and $E-m>0$ then $E^2-m^2<0$. Inequality $E-m>0$ leads to $E>m>0$, but $E^2-m^2<0$ leads to $E<-m<0$, which is a contradiction. Then if $V\le 0$ and $E-m<0$ we should have $E^2-m^2>0$. Both inequalities $E-m<0$ and $E^2-m^2>0$ lead to $E<-m$. 

Now we assume that $\lim\limits_{x\to\infty}V=0$, then system (\ref{d1l})--(\ref{d1r}) asymptotically becomes
\begin{subnumcases}{}
\varphi_1'=-(E+m)\varphi_2,\\
\varphi_2'=\ph(E-m)\varphi_1.
\end{subnumcases}
If $\varphi_1\ge 0$ before vanishing, then $\varphi_1'\le 0$ and, using $E<-m$, above system yields $\varphi_2\le 0$ and $\varphi_2'\le 0$ near infinity, which is the contradiction. Assumption $\lim\limits_{x\to\infty}V=-\infty$ leads to
\begin{subnumcases}{}
\varphi_1'=2V\varphi_2,\\
\varphi_2'=(E-m)\varphi_1.
\end{subnumcases} 
Now if $\varphi_1\ge 0$ and $\varphi_1'\le 0$ we have $\varphi_2\ge 0$ and $\varphi_2'\le 0$, which means that $\varphi_2$ approaches zero with positive sign. Finally we conclude that if $S=-V$ and $V\le 0$ then $E<-m$ and $\lim\limits_{x\to\infty}V=-\infty$, which corresponds to {\it (2)}. Following the same path, the case $s=1$ and the remaining classes of potential and corresponding inequalities can be established.

%%%%%%%%%%%%%%%%%%%%%%%%%%%%%%%%%%%%%%%%%%%%%%%%%%%%%%%%%%%%%%%%%%%%%%%%%%%%%%%%%%%%%%%%
\subsection{Refined comparison theorems}
%%%%%%%%%%%%%%%%%%%%%%%%%%%%%%%%%%%%%%%%%%%%%%%%%%%%%%%%%%%%%%%%%%%%%%%%%%%%%%%%%%%%%%%%
Suppose that $\{\varphi_1(x),\ \varphi_2(x)\}$ is a solution of the Dirac coupled equations (\ref{d1l})--(\ref{d1r}). Since the potential $V$ is an even function, it follows from (\ref{d1l})--(\ref{d1r}) that $\{\varphi_1(-x),\ -\varphi_2(-x)\}$ and $\{-\varphi_1(-x),\ \varphi_2(-x)\}$ are also solutions of (\ref{d1l})--(\ref{d1r}). Thus $\varphi_1$ and $\varphi_2$ have definite and opposite parities, i.\,e. if $\varphi_1$ is even then $\varphi_2$ is odd and {\it vice versa}. Therefore, because of the symmetry of the wave functions, we shall consider only the positive half axis $x\ge 0$.  

Now we prove the lemma which characterizes the behaviour of the one dimensional Dirac wave functions in the ground state: 

\medskip

\noindent{\bf Lemma 1:} ~~{\it In the ground state the upper $\varphi_1$ and lower $\varphi_2$ components of the Dirac spinor are monotone in the spin and pseudo--spin symmetric cases respectively.} 

\medskip

\noindent{\bf Proof:} In the $s=-1$ case equation (\ref{d1r}) becomes 
\begin{equation}\label{S-V4}
\varphi_{2}'=(E-m)\varphi_{1}.
\end{equation}
Since in the ground state $\varphi_1$ has constant sign, the function $\varphi_{2}'$ has constant sign as well, which result ends the proof. The case $s=1$, for which the roles of $\varphi_1$ and $\varphi_2$ are interchanged, can be similarly proved.

\hfill $\Box$\\

For example, consider the $s=-1$ case with potential $V$ satisfying {\it(2)}. We are looking for the ground state. Thus without loss of generality, we put $\varphi_1\ge 0$ on $[0,\ \infty)$. Then equation (\ref{S-V4}) yields $\varphi_{2}'\le 0$, so $\varphi_2$ has to be even and nonnegative. Consequently $\varphi_1$ is odd, so $\varphi_1'$ must change its sign from positive to negative. In order to guarantee such behaviour of $\varphi_1$, the potential $V$ has to be smaller then $E+m$ near the origin and then dominate the term $E+m$ at infinity: this is true since $V(0)=0$ and $\lim\limits_{|x|\to\infty}V=-\infty$.

Now we refine the basic comparison theorem which follows from relation (\ref{VS}). 

\medskip

\noindent{\bf Theorem 1:} ~~{\it The potential $V$  belongs to one of the classes (1)--(3) and has area, $S=sV$, and
\begin{equation}\label{th3}
g(x)=\int_0^x (V_b(t)-V_a(t))dt, \quad x\in[0,\ \infty).
\end{equation}
Then if $g\ge 0$, the eigenvalues are ordered, i.\,e. $E_a\le E_b$.} 

\medskip

\noindent{\bf Proof:} ~~We prove the theorem for the pseudo--spin symmetric case, i.\,e. $s=-1$; for the other case the proof is essentially the same. We integrate the right side of (\ref{VS}) by parts to obtain
\begin{equation*}
2\int_0^\infty (V_b-V_a)\varphi_{2a}\varphi_{2b}dx=\varphi_{2a}\varphi_{2b}g|_0^\infty -2\int_0^\infty g\left(\varphi_{2a}\varphi_{2b}\right)'dx,
\end{equation*}
where $g$ is defined by (\ref{th3}). Since $g(0)=0$ and $\lim\limits_{x\to\infty}\varphi_2=0$, relation (\ref{VS}) becomes
\begin{equation*}
(E_b-E_a)\int_0^\infty (\varphi_{1a}\varphi_{1b}+\varphi_{2a}\varphi_{2b})dx=
-2\int_0^\infty g\left(\varphi_{2a}\varphi_{2b}\right)'dx,
\end{equation*}
According to Lemma 1, $\varphi_2$ is monotone and, since it is also square integrable, it follows that the functions $\varphi_2$ and $\varphi_2'$ have different signs in the ground state, i.\,e. if $\varphi_2\ge 0$ then $\varphi_2'\le 0$ on $[0,\ \infty)$ and {\it vice versa}. Thus the derivative of the product satisfies $\left(\varphi_{2a}\varphi_{2b}\right)'\le 0$. Finally, if $g\ge 0$, it follows from the above expression that $E_a\le E_b$.

\hfill $\Box$\\

If we know more details of the interlacing relations of the comparison potentials, we can state a corollary of the above theorem which is easier to apply on practice:

\medskip

\noindent{\bf Corollary 1:} ~~{\it Let the comparison potentials belong to one of the classes (1)--(3). If the potentials cross over once, say at $x_1$, $V_a\le V_b$ for $x\in [0,\ x_1]$, and
\begin{equation*}
g(\infty)=\int_0^\infty (V_b-V_a)dx,  
\end{equation*}
or if the potentials cross over twice, say at $x_1$ and $x_2$, $x_1<x_2$, $V_a\le V_b$ for $x\in [0,\ x_1]$, and
\begin{equation*}
g(x_2)=\int_0^{x_2} (V_b-V_a) dx. 
\end{equation*}
Then if $g(\infty)\ge 0$ and $g(x_2)\ge 0$ it follows that $g(x)\ge 0$ and the eigenvalues are ordered, i.\,e. $E_a\le E_b$.} 

\medskip

We can extend Corollary 1 to the case of $n$ intersections, $n=1,\ 2,\ 3,\ \ldots$, say at points $x_1,\ x_2,\ x_3,\ \ldots$. As before we suppose that $V_a\le V_b$ on the first interval $x\in[0,\ x_1]$. Then we assume that the sequence $\int_{x_i}^{x_{i+1}}|V_b-V_a|dx$, $i=1,\ 2,\ 3,\ \ldots,\ n$, of absolute areas is nonincreasing (if $n$ is odd then $\int_{x_{n-1}}^{x_n}|V_b-V_a|dx\ge\int_{x_n}^\infty|V_b-V_a|dx$ ), this leads to $g\ge 0$ on  $x\in[0,\ \infty)$ thus, according to the first theorem, $E_a\le E_b$. 

Now we state and give proof of the second refined comparison theorem. Where the difference $V_b-V_a$ is multiplied by upper $\varphi_1$ or lower $\varphi_2$ component of the Dirac spinor.  

\medskip

\noindent{\bf Theorem 2:} ~~{\it The potential $V$  belongs to one of the classes (1)--(3) and has $\varphi_l$--weighted area, $S=sV$, and 
\begin{equation}\label{th4}
p(x)=\int_0^x (V_b(t)-V_a(t))|\varphi_{l}(t)|dt, \quad x\in [0,\ \infty).
\end{equation}
Then if $p\ge 0$, the eigenvalues are ordered, i.\,e. $E_a\le E_b$, where $\varphi_l=\varphi_{1i}$ if $s=1$ and $\varphi_l=\varphi_{2i}$ if $s=-1$, $i=a$ {\bf or} $b$.} 

\medskip

\noindent{\bf Proof:} ~~We prove the theorem for the spin symmetric case and assume that the upper component of the Dirac spinor is known, so $s=1$ and $\varphi_l=\varphi_{1i}$; for the other case the proof is essentially the same. The right side of (\ref{VS}) after integration by parts becomes
\begin{equation*}
2\int_0^\infty (V_b-V_a)\varphi_{1a}\varphi_{1b}dx=\varphi_{1b}p|_0^\infty
-2\int_0^\infty p\left(\varphi_{1b}\right)'dx,
\end{equation*}
where $p$ is defined by (\ref{th4}) for $\varphi_{1i}=\varphi_{1a}$. The expression $\varphi_{1b}p|_0^\infty=0$, because $p(0)=0$ and $\lim\limits_{x\to\infty}\varphi_{1}=0$. Then relation (\ref{VS}) takes the form
\begin{equation*}
(E_b-E_a)\int_0^\infty (\varphi_{1a}\varphi_{1b}+\varphi_{2a}\varphi_{2b})dx=
-2\int_0^\infty p\left(\varphi_{1b}\right)'dx.
\end{equation*}
Functions $\varphi_1$ and $\varphi_1'$ have different signs thus $p\left(\varphi_{1b}\right)'\le 0$ and we conclude $E_a\le E_b$, which inequality establishes the theorem.

\hfill $\Box$\\

The wave functions vanish at infinity, thus the potential difference might be bigger in the second theorem than in the first one and still lead to $E_a\le E_b$. As before we can formulate simpler sufficient condition for spectral ordering if more detailed potential behaviour is known:

\medskip

\noindent{\bf Corollary 2:} ~~{\it Let the comparison potentials belong to one of the classes (1)--(3). If the potentials cross over once, say at $x_1$, $V_a\le V_b$ for $x\in [0,\ x_1]$, and
\begin{equation*}
p(\infty)=\int_0^\infty (V_b-V_a)|\varphi_{l}|dx,
\end{equation*}
or if the potentials cross over twice, say at $x_1$ and $x_2$, $x_1<x_2$, $V_a\le V_b$ for $x\in [0,\ x_1]$, and
\begin{equation*}
p(x_2)=\int_0^{x_2} (V_b-V_a)|\varphi_{l}|dx.
\end{equation*}
Then if $p(\infty)\ge 0$ or $p(x_2)\ge 0$ it follows that $p(x)\ge 0$ and the eigenvalies are ordered, i.\,e. $E_a\le E_b$, where $\varphi_l=\varphi_{1i}$ if $s=1$ and $\varphi_l=\varphi_{2i}$ if $s=-1$, $i=a$ {\bf or} $b$.} 

\medskip

Corollary 2 can also be generalized for the case of $n$ intersections: if $V_a\le V_b$ on $x\in[0,\ x_1]$ and the sequence $\int_{x_i}^{x_{i+1}}|(V_b-V_a)\varphi_{l}|dx$, $i=1,\ 2,\ 3,\ \ldots,\ n$ and $\varphi_l=\varphi_{1i}$ if $s=1$ and $\varphi_l=\varphi_{2i}$ if $s=-1$, $i=a$ or $b$, is nonincreasing (and, if $n$ is odd, $\int_{x_{n-1}}^{x_n}|(V_b-V_a)\varphi_{l}|dx\ge\int_{x_n}^\infty|(V_b-
V_a)\varphi_{l}|dx$), then $p\ge 0$ on $x\in[0,\ \infty)$, so, according to Theorem 2, we conclude $E_a\le E_b$.

%%%%%%%%%%%%%%%%%%%%%%%%%%%%%%%%%%%%%%%%%%%%%%%%%%%%%%%%%%%%%%%%%%%%%%%%%%%%%%%%%%%%%%%%
%%%%%%%%%%%%%%%%%%%%%%%%%%%%%%%%%%%%%%%%%%%%%%%%%%%%%%%%%%%%%%%%%%%%%%%%%%%%%%%%%%%%%%%%
\subsection{An Example}
%%%%%%%%%%%%%%%%%%%%%%%%%%%%%%%%%%%%%%%%%%%%%%%%%%%%%%%%%%%%%%%%%%%%%%%%%%%%%%%%%%%%%%%%
%%%%%%%%%%%%%%%%%%%%%%%%%%%%%%%%%%%%%%%%%%%%%%%%%%%%%%%%%%%%%%%%%%%%%%%%%%%%%%%%%%%%%%%%
In this section as an example we consider the extension of Corollary 1 to the case of $n$ intersections in the spin symmetric case. We take the harmonic oscillator $V_a$ and a modified harmonic oscillator $V_b$ as our comparison potentials:
\begin{equation*}
V_a=ax^2 \qquad \text{and} \qquad 
V_b=bx^2\left(1+\frac{\sin(x^3+\beta)}{x^3+\beta}\right).
\end{equation*}
Both comparison potentials satisfy {\it(2)} for $s=1$. If $a=b$ the substitution $z=x^3+\beta$ transforms the integral (\ref{th3}) into
\begin{equation*}
\int_0^\infty (V_b-V_a)dt=\frac{b}{3}\int_\beta^\infty\frac{\sin z}{z}dz. 
\end{equation*}
Choosing $\beta=1.64$, and calculating numerical values, we find that the first area is bigger then the second one:
\begin{equation*}
\int_\beta^\pi \frac{|\sin z|}{z}dz =0.43810>\int_\pi^{2\pi} \frac{|\sin z|}{z}dz=0.43379.
\end{equation*} 
The $\sin z$ is a periodic function, thus $|\sin x|=|\sin y|$ where $x\in[(k-1)\pi,\ k\pi]$ and $y=x+\pi$, $k=3,\ 4,\ 5,\ \ldots$, then it is clear that 
\begin{equation*}
\int_{(k-1)\pi}^{k\pi} \frac{|\sin z|}{z}dz >\int_{k\pi}^{(k+1)\pi} \frac{|\sin z|}{z}dz.
\end{equation*}
Therefore 
\begin{equation*}
\int_0^\infty (V_b-V_a)dt\ge 0,
\end{equation*}
because successive positive and negative areas of the integrand do not increase in absolute value. Thus $g>0$ and by Theorem 1 we have $E_a\le E_b$.  
This prediction is verified by accurate numerical calculations:
for $a=b=0.5$, $\beta=1.64$, and $m=1.2$ the comparison potentials intersect at infinitely many points (see Figure 1) and numerical eigenvalues are $E_a=1.77935\le E_b=1.85470$.
\begin{figure}
\centering{\includegraphics[height=13cm,width=7cm,angle=-90]{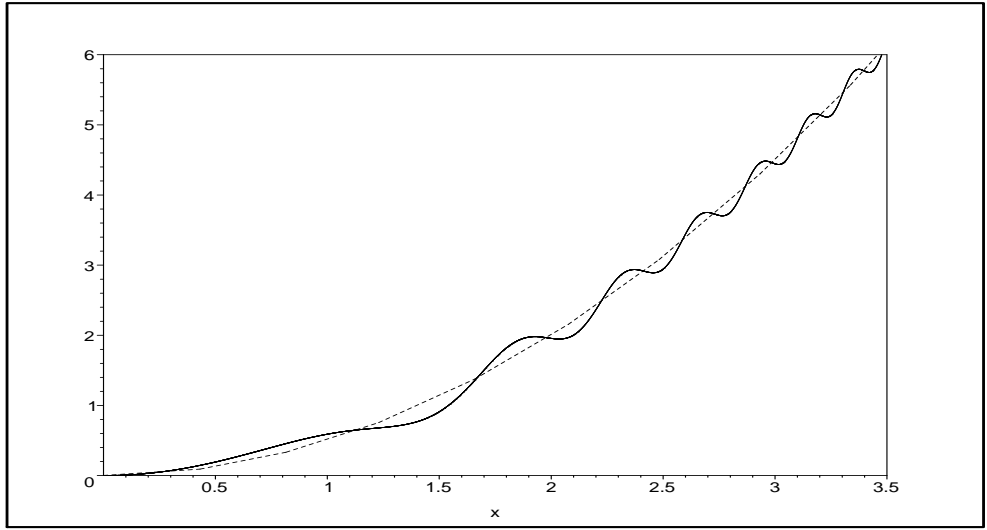}}
\caption{Potential $V_a$ dashed lines and $V_b$ full line.}
\end{figure}

%%%%%%%%%%%%%%%%%%%%%%%%%%%%%%%%%%%%%%%%%%%%%%%%%%%%%%%%%%%%%%%%%%%%%%%%%%%%%%%%%%%%%%%%
%%%%%%%%%%%%%%%%%%%%%%%%%%%%%%%%%%%%%%%%%%%%%%%%%%%%%%%%%%%%%%%%%%%%%%%%%%%%%%%%%%%%%%%%
\section{The $d$-dimensional case}
%%%%%%%%%%%%%%%%%%%%%%%%%%%%%%%%%%%%%%%%%%%%%%%%%%%%%%%%%%%%%%%%%%%%%%%%%%%%%%%%%%%%%%%%
%%%%%%%%%%%%%%%%%%%%%%%%%%%%%%%%%%%%%%%%%%%%%%%%%%%%%%%%%%%%%%%%%%%%%%%%%%%%%%%%%%%%%%%%
%%%%%%%%%%%%%%%%%%%%%%%%%%%%%%%%%%%%%%%%%%%%%%%%%%%%%%%%%%%%%%%%%%%%%%%%%%%%%%%%%%%%%%%%
%%%%%%%%%%%%%%%%%%%%%%%%%%%%%%%%%%%%%%%%%%%%%%%%%%%%%%%%%%%%%%%%%%%%%%%%%%%%%%%%%%%%%%%%
\subsection{The Dirac equation in $d$ dimensions}
%%%%%%%%%%%%%%%%%%%%%%%%%%%%%%%%%%%%%%%%%%%%%%%%%%%%%%%%%%%%%%%%%%%%%%%%%%%%%%%%%%%%%%%%
%%%%%%%%%%%%%%%%%%%%%%%%%%%%%%%%%%%%%%%%%%%%%%%%%%%%%%%%%%%%%%%%%%%%%%%%%%%%%%%%%%%%%%%%
The Dirac equation in $d>1$ dimensions is given by \cite{jiang}
\begin{equation*}
i\frac{\partial \Psi}{\partial t} =H\Psi,\quad {\rm where}\quad  H=\sum_{s=1}^{d}{\alpha_{s}p_{s}} + (m+S)\beta+V,
\end{equation*}
where we use natural units $\hbar=c=1$, $m$ is the mass of the particle, the functions $V$ and $S$ are spherically symmetric vector and scalar potentials, and $\{\alpha_{s}\}$ and $\beta$  are Dirac matrices, which satisfy anti--commutation relations; the identity matrix is implied after the potential $V$. The above equation can be written as the following system of two first--order differential equations \cite{jiang, agboola, salazar, yasuk}
\begin{subnumcases}{}
\label{Sl1}
\psi_1'=(m+E+S-V)\psi_2-\frac{k_d}{r}\psi_1,\\
\label{Sr1}
\psi_2'=(m-E+S+V)\psi_1+\frac{k_d}{r}\psi_2,
\end{subnumcases}
where $\psi_1$ and $\psi_2$ are radial wave functions, $r = \|\mb{r}\|$, prime $'$ denotes the derivative with respect to $r$, $k_d=\tau\left(j+\frac{d-2}{2}\right)$, $\tau = \pm 1$, and $j=1/2$, $3/2$, $5/2$, $\ldots$. We assume that the potentials $V$ and $S$ are such that there is an energy eigenvalue $E$ and that equations (\ref{Sl1})--(\ref{Sr1}) are the eigenequations for the corresponding pair of radial eigenstates. For $d > 1,$ the wave functions vanish at $r = 0$, and for bound states they obey the normalization condition 
\begin{equation*}
(\psi_1,\psi_1)+(\psi_2,\psi_2)=\int\limits_0^{\infty}(\psi_1^2 + \psi_2^2)dr = 1.
\end{equation*}

As in one dimension, we now compare the system (\ref{Sl1})--(\ref{Sr1}) for the eigenvalues respectively $E_a$ and $E_b$: 
\begin{subnumcases}{}
\label{Sl1a}
\psi_1'=(m+E_a+S_a-V_a)\psi_{2a}-\frac{k_d}{r}\psi_{1a},\\
\label{Sr1a}
\psi_2'=(m-E_a+S_a+V_a)\psi_{1a}+\frac{k_d}{r}\psi_{2a},
\end{subnumcases}
and
\begin{subnumcases}{}
\label{Sl1b}
\psi_1'=(m+E_b+S_b-V_b)\psi_{2b}-\frac{k_d}{r}\psi_{1b},\\
\label{Sr1b}
\psi_2'=(m-E_b+S_b+V_b)\psi_{1b}+\frac{k_d}{r}\psi_{2b}.
\end{subnumcases}
Then we form the following combination of the equations: (\ref{Sl1a})$\psi_{2b}-$ (\ref{Sr1a})$\psi_{1b}-$ (\ref{Sl1b})$\psi_{2a}+$ (\ref{Sr1b})$\psi_{1a}$, which, after integration and some simplifications, takes the form
\begin{equation}\label{expr6}
(E_b - E_a)\int_0^\infty (\psi_{1a}\psi_{1b} + \psi_{2a}\psi_{2b})dr=\int_0^\infty\left[(V_b-V_a-S_a+S_b)\psi_{1a}\psi_{1b}+
(V_b-V_a+S_a-S_b)\psi_{2a}\psi_{2b}\right]dr.
\end{equation}
By introducing the parameter $s$, we can combine the spin and pseudo--spin symmetric cases, i.\,e. $S=sV$ where $s=1$ if $S=V$ and $s=-1$ if $S=-V$. Then the above expression for the $S=sV$ case becomes 
\begin{equation}\label{expr7}
(E_b - E_a)\int_0^\infty (\psi_{1a}\psi_{1b} + \psi_{2a}\psi_{2b})dr=2\int_0^\infty(V_b-V_a)\psi_{qa}\psi_{qb}dr,
\end{equation}
where $q=1$ if $s=1$ and $q=2$ if $s=-1$. If the wave functions are nodeless, i.\,e. have constant sign on $[0,\ \infty)$, and the potentials are ordered, say $V_a\le V_b$, then the integrands of (\ref{expr7}) have constant sign and $E_a\le E_b$, which is equivalent to the usual comparison theorem. We shall refine that theorem later, as in the one-dimensional case. For example, we may replace $V_a\le V_b$ by the weaker condition $\int_0^r V_b(t)t^{-2sk_d} dt\ge \int_0^r V_a(t)t^{-2sk_d} dt$ for some cases. We shall consider theorems for specific classes of potentials in section C. below.

Now, if two comparison scalar potentials $S_a$ and $S_b$ are equal but the vector potentials $V_a$ and $V_b$ are different, i.\,e. $S_a=S_b$ and $V_a\ne V_b$, the relation (\ref{expr6}) can be rewritten as 
\begin{equation}\label{expr8}
(E_b - E_a)\int_0^\infty (\psi_{1a}\psi_{1b} + \psi_{2a}\psi_{2b})dr=\int_0^\infty(V_b-V_a)(\psi_{1a}\psi_{1b}+
\psi_{2a}\psi_{2b})dr.
\end{equation}
Then the following comparison theorem immediately follows:

\medskip

\noindent{\bf Theorem 3:} ~~{\it If $S_a=S_b$ and $V_a\le V_b$, then $E_a\le E_b$.} 

\medskip

As an example we consider the Coulomb potential $S_a=S_b=-\cfrac{s}{r}$, with $s=0.7$. For the vector potentials we choose the soft--core potential \cite{softcore_1, softcore_2} $V_a=-\cfrac{\alpha}{\left(r^q+a^q\right)^{1/q}}$ and sech--squared potential \cite{sechsquared_4, sechsquared_2, sechsquared_1, sechsquared_3} $V_b=-\cfrac{4\beta}{\left(e^{br}+e^{-br}\right)^2}$. If $\alpha=0.8$, $a=1.6$, $q=3$, $\beta=0.5$, and $b=0.31$ the potentials are ordered $V_a\le V_b$. Then, by Theorem 3, we conclude $E_a\le E_b$, which is verified by accurate numerical eigenvalues $E_a=0.77260\le E_b=0.81648$ for $m=1$, $\tau=-1$, $d=5$, and $j=1/2$.

We note that expression (\ref{expr8}) is exactly the same as (11) from the recent work \cite{Dir}. Therefore Theorem 3 can be refined in the same manner and corresponding corollaries can be derived {\it mutatis mutandis}. 

We also note that one can derive similar theorem in one dimension. That is to say, we can obtain the expression
\\
$(E_b - E_a)\int_0^\infty (\varphi_{1a}\varphi_{1b} + \varphi_{2a}\varphi_{2b})dx=\int_0^\infty(V_b-V_a)(\varphi_{1a}\varphi_{1b}+
\varphi_{2a}\varphi_{2b})dx$ from (\ref{expr1}) and conclude that if $S_a=S_b$ and $V_a\le V_b$, then $E_a\le E_b$.

%%%%%%%%%%%%%%%%%%%%%%%%%%%%%%%%%%%%%%%%%%%%%%%%%%%%%%%%%%%%%%%%%%%%%%%%%%%%%%%%%%%%%%%%
\subsection{Classes of potentials}
%%%%%%%%%%%%%%%%%%%%%%%%%%%%%%%%%%%%%%%%%%%%%%%%%%%%%%%%%%%%%%%%%%%%%%%%%%%%%%%%%%%%%%%%
Here we characterize the relationship between the eigenvalue $E$ and mass of the particle $m$ depending on the type of the potential $V$. As in one dimension, equations (\ref{Sl1})--(\ref{Sr1}) can be written in a Schr\"{o}dinger--like form
\begin{equation}\label{Schd}
-\psi '' +\left(\cfrac{k_d(k_d+s)}{r^2}+2(E+sm)V\right)\psi=-(m^2-E^2)\psi,
\end{equation}
where $\psi=\psi_1$ if $s=1$ and $\psi=\psi_2$ if $s=-1$. We shall consider the following three classes of potential:
\\
 
{\it $(i)$ $V$ is finite near infinity and\\

\hspace{3cm} (1) $sV\le 0$ and $\lim\limits_{r\to\infty}V=0$. This implies $-m<E<m$;\\

$(ii)$ $V$ is unbounded near infinity and\\

\hspace{3cm} (2) $sV\ge 0$ and $\lim\limits_{r\to\infty}V=\ph s\infty$. This implies $sE>\ph m$\\

or\\

\hspace{3cm} (3) $sV\le 0$ and $\lim\limits_{r\to\infty}V=-s\infty$. This implies $sE<-m$.}\\

Following a similar path as in one dimension, one can verify that the above classes of potentials and relations between energy $E$ and mass $m$ are valid for the system of the Dirac coupled equations (\ref{Sl1})--(\ref{Sr1}) under spin and pseudo--spin symmetry.

%%%%%%%%%%%%%%%%%%%%%%%%%%%%%%%%%%%%%%%%%%%%%%%%%%%%%%%%%%%%%%%%%%%%%%%%%%%%%%%%%%%%%%%%
%%%%%%%%%%%%%%%%%%%%%%%%%%%%%%%%%%%%%%%%%%%%%%%%%%%%%%%%%%%%%%%%%%%%%%%%%%%%%%%%%%%%%%%%
\subsection{Refined comparison theorems in $d$ dimensions}
%%%%%%%%%%%%%%%%%%%%%%%%%%%%%%%%%%%%%%%%%%%%%%%%%%%%%%%%%%%%%%%%%%%%%%%%%%%%%%%%%%%%%%%%
%%%%%%%%%%%%%%%%%%%%%%%%%%%%%%%%%%%%%%%%%%%%%%%%%%%%%%%%%%%%%%%%%%%%%%%%%%%%%%%%%%%%%%%%
In that section we refine relativistic comparison theorems in a way that the graphs of the potentials can crossover in a controlled manner with the preservation of spectral ordering. Our establishment of refined comparison theorems requires monotone behaviour of the wave function and consequently a constant sign of its derivative. But bound state wave functions are zero at the origin and vanish at infinity. Thus even if the wave function has constant sign, its derivative changes sign. The following lemma helps us to allow for this.

\medskip

\noindent{\bf Lemma 2:} ~~{\it At the bottom of an angular--momentum subspace labelled by $j$, the functions $\psi_1 r^{k_d}$ and $\psi_2 r^{-k_d}$ are monotone in the spin and pseudo--spin symmetric cases respectively.} 

\medskip

\noindent{\bf Proof:} In the case $s=1$, using (\ref{Sl1})--(\ref{Sr1}), we find
\begin{equation*}
\left(\psi_1 r^{k_d}\right)'=(m+E)\psi_2r^{k_d}.
\end{equation*}
Clearly $\left(\psi_1 r^{k_d}\right)'$ has constant sign since $m+E$ is constant and $\psi_2$ is either nonpositive or nonnegative. The case $s=-1$ can be proven similarly.

\hfill $\Box$\\

As in thee one--dimensional case, we need to know some characteristics of the nodeless state of the Dirac coupled equations (\ref{Sl1})--(\ref{Sr1}). For example, consider the case ${\it (1)}$ with $s=1$:  according to the previous section, the potential $V\le 0$, $\lim\limits_{r\to\infty}V=0$, and $-m<E<m$. The system (\ref{Sl1})--(\ref{Sr1}) then takes the following form
\begin{subnumcases}{}
\label{Sl2}
\psi_1'=(m+E)\psi_2-\frac{k_d}{r}\psi_1,\\
\label{Sr2}
\psi_2'=(m-E+2V)\psi_1+\frac{k_d}{r}\psi_2.
\end{subnumcases}
Asymptotically near infinity the above equations become, 
\begin{subnumcases}{}
\label{Sl5}
\psi_1'=(m+E)\psi_2,\\
\label{Sr5}
\psi_2'=(m-E)\psi_1.
\end{subnumcases}   
The components $\psi_1$ and $\psi_2$ of the Dirac spinor vanish at infinity. Suppose that $\psi_1\ge 0$ before vanishing, then $\psi_1'\le 0$ and it follows from the system above that $\psi_2\le 0$ and $\psi_2'\ge 0$. The assumption $\psi_1\le 0$ before vanishing, leads to $\psi_2\ge 0$ and $\psi_2'\le 0$. Consequently $\psi_1$ and $\psi_2$ must vanish with different signs.

Near the origin if $k_d>0$, we set $\psi_1\ge 0$ so $\psi_1'\ge 0$, then equation (\ref{Sl2}) leads to $\psi_2\ge 0$. The assumption $\psi_1\le 0$ would give $\psi_2\le 0$. The quantity $m-E+2V$ has to change sign to guarantee the necessary behaviour of $\psi_2$, i.\,e. increasing then decreasing if it is nonnegative or decreasing then increasing if it is nonpositive. Since $\lim\limits_{r\to\infty}V=0$, then $\lim\limits_{r\to\infty}(m-E+2V)=m-E>0$, so $\lim\limits_{r\to 0^+}(m-E+2V)<0$, thus $m-E+2V$ changes sign exactly once from negative to positive (more details can be found in \cite{nod}). Then for $k_d<0$ equation (\ref{Sr2}) leads to $\psi_1\le 0$ if $\psi_2\ge 0$ and {\it vice versa}. Hence, if $k_d>0$ both wave function components start at the origin with the same sign, but at infinity they must have different signs: thus one of the wave function components will have at least one node in the lowest state (Figure 2, left graph). When $k_d<0$, $\psi_1$ and $\psi_2$ start with different signs and then vanish with different signs: thus neither of them has a node in the ground state (Figure 2, right graph). 
\begin{figure}[ht]
\begin{center}$
\begin{array}{cc}
\includegraphics[width=2.5in, angle=-90]{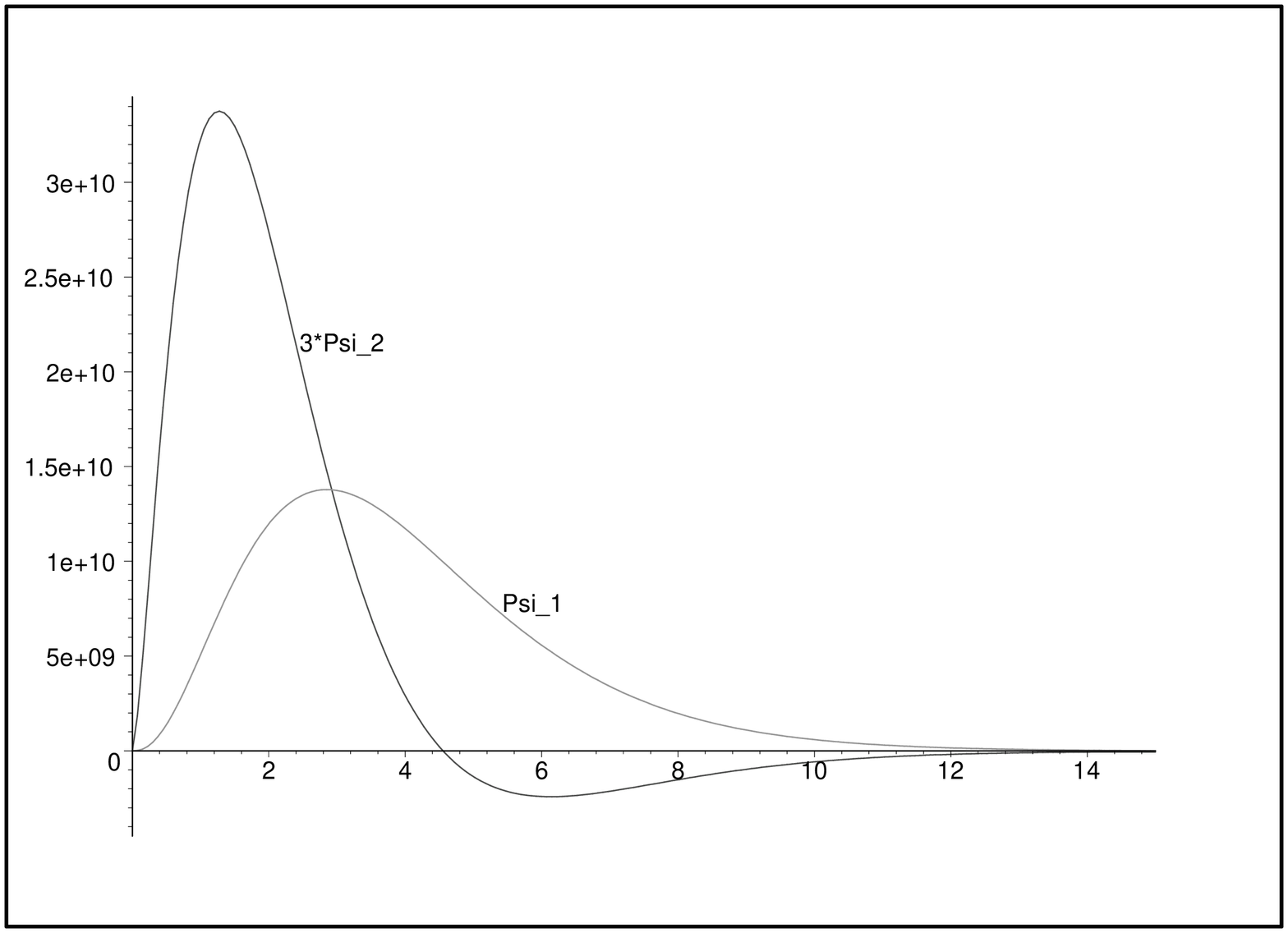} & 
\includegraphics[width=2.5in, angle=-90]{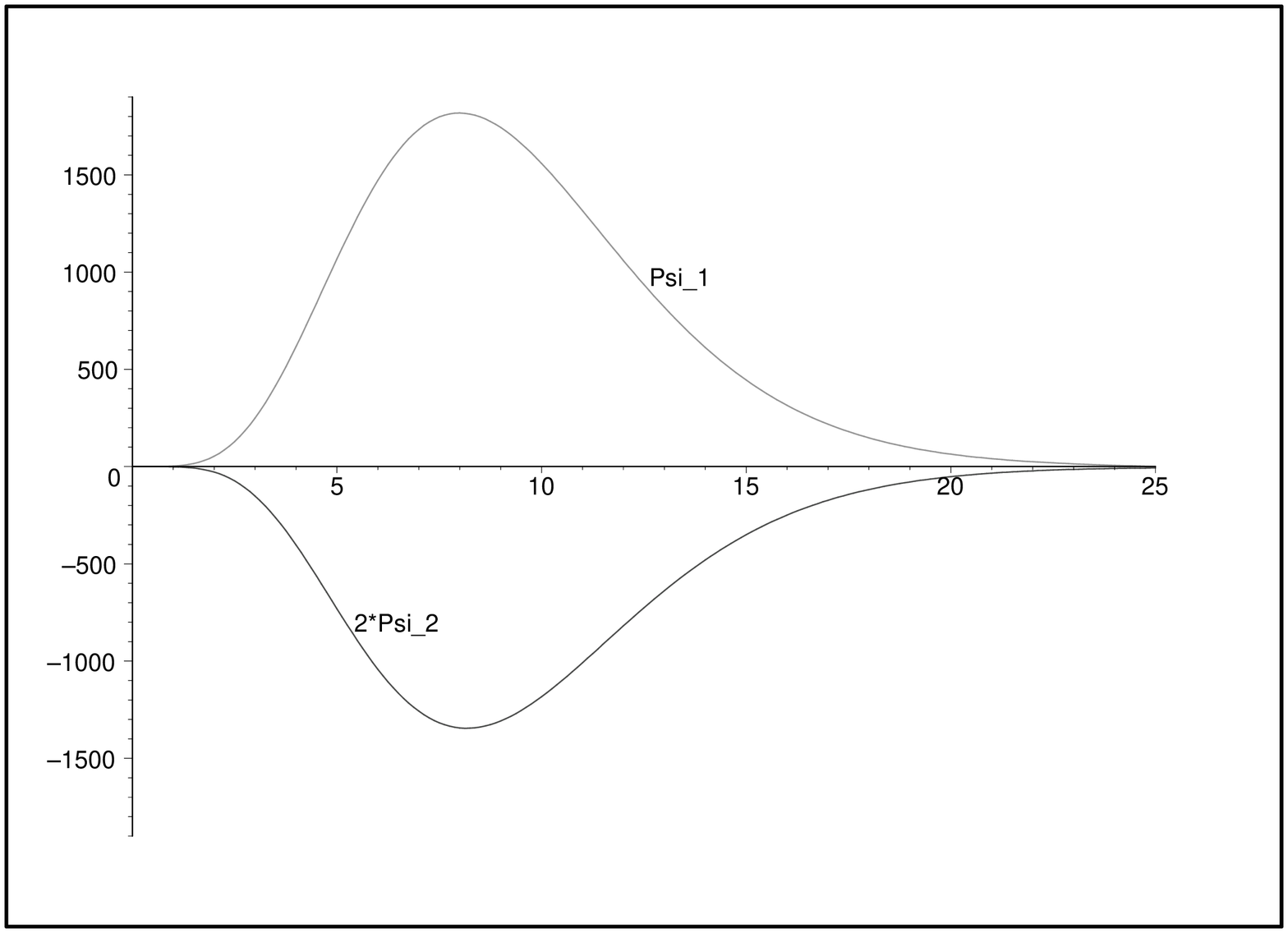}
\end{array}$
\end{center}
\caption{Ground state of the Dirac coupled equations (\ref{Sl1})--(\ref{Sr1}) in the spin--symmetric case, $V=S$, for the cut--off Coulomb potential \cite{cutoff_2, Barton, cutoff_1} $V=-\cfrac{v}{r+a}$. Left graph: $\tau=1$, $m=1$, $d=4$, $j=1/2$, $v=1.5$, $a=0.01$, and $E=0.47399$. Right graph: $\tau=-1$, $m=1$, $d=7$, $j=5/2$, $v=2.5$, $a=1.2$, and $E=0.69329$. }
\end{figure}
Similarly analysing the case $s=-1$ and other types of potential, we get that $\psi_1$ and $\psi_2$ have no nodes if $k_d>0$. Finally, we infer: {\it the Dirac radial wave functions $\psi_1$ and $\psi_2$, which satisfy (\ref{Sl1})--(\ref{Sr1}), are node free in the case $S=sV$ if $sk_d<0$}. We note that Alberto {\it et. al.} in the recent work \cite{Al} derived general result: $n_1=n_2$ if $sk_d<0$, where $n_1$ and $n_2$ are the numbers of nodes of $\psi_1$ and $\psi_2$ respectively. Now, using this result, we state and prove the refined comparison theorem. 

\newpage

\noindent{\bf Theorem 4:} ~~{\it The potential $V$ belongs to one of the classes ${\it (1)}$--${\it (3)}$ and has $r^{-2sk_d}$--weighted area, $S=sV$, $sk_d<0$, and
\begin{equation}\label{th8}
\rho(r)=\int_0^r(V_b(t)-V_a(t))t^{-2sk_d}dt, \quad r\in [0,\ \infty).
\end{equation}
Then if $\rho\ge 0$, the eigenvalues are ordered, i.\,e. $E_a\le E_b$.}

\medskip

\noindent{\bf Proof:} We prove the theorem for the spin symmetric case, i.\,e. $s=1$; for the other case the proof is similar. Let us integrate by parts the right side of (\ref{expr7}) in the following way 
\begin{equation*}
\int_0^\infty(V_b-V_a)\psi_{1a}\psi_{1b}dr=
\left.\psi_{1a}\psi_{1b}\rho r^{2k_d}\right|_0^\infty-
\int_0^\infty\rho\left(\psi_{1a}\psi_{1b}r^{2k_d}\right)'dr,
\end{equation*}
where $\rho$ is defined by (\ref{th8}). Since $\rho(0)=0$ and $\lim\limits_{r\to\infty}\psi_{1}=0$, relation (\ref{expr7}) becomes
\begin{equation}\label{eq35}
(E_b - E_a)\int_0^\infty (\psi_{1a}\psi_{1b} + \psi_{2a}\psi_{2b})dr=
-\int_0^\infty\rho\left(\psi_{1a}\psi_{1b}r^{2k_d}\right)'dr.
\end{equation}
Since $\psi_1$ vanishes at infinity, the function $\psi_1 r^{k_d}$ vanishes as well. Thus, according to Lemma 2, the functions $\left(\psi_1 r^{k_d}\right)'$ and $\psi_1 r^{k_d}$ have different signs, which leads to $\left(\psi_{1a}\psi_{1b}r^{2k_d}\right)'\le 0$. Then it follows from expression (\ref{eq35}) that the nonnegativity of $\rho$ and the nodeless form of the wave functions result in $E_a\le E_b$.

\hfill $\Box$

As in the one--dimensional case, if we know more details concerning the behaviour of the comparison potentials, we can state simpler sufficient conditions:

\medskip

\noindent{\bf Corollary 4:} ~~{\it Let the comparison potentials belong to one of the classes ${\it (1)}$--${\it (3)}$. If the potentials cross over once, say at $r_1$, $V_a\le V_b$ for $r\in [0,\ r_1]$, and
\begin{equation*}
\rho(\infty)=\int_0^\infty (V_b-V_a)r^{-2sk_d}dr,  
\end{equation*}
or if the potentials cross over twice, say at $r_1$ and $r_2$, $r_1<r_2$, $V_a\le V_b$ for $r\in [0,\ r_1]$, and
\begin{equation*}
\rho(r_2)=\int_0^{r_2} (V_b-V_a)^{-2sk_d} dr. 
\end{equation*}
Then if $\rho(\infty)\ge 0$ and $g(x_2)\ge 0$ it follows that $\rho\ge 0$ and the eigenvalues are ordered, i.\,e. $E_a\le E_b$.} 

\medskip

We can extend the above corollary in the following way: assume that comparison potentials have $n$ intersections, $n=1,\ 2,\ 3,\ \ldots$, and $V_a\le V_b$ on $r\in[0,\ r_1]$. Also assume that $\int_{r_i}^{r_{i+1}}|V_b-V_a|r^{-2sk_d}dr$, $i=1,\ 2,\ 3,\ \ldots,\ n$ (if $n$ is odd then $\int_{r_{n-1}}^{r_n}|V_b-V_a|r^{-2sk_d}dr\ge\int_{r_n}^\infty|V_b-V_a|r^{-2sk_d}dr$ ), hence $\rho(r)\ge 0$ for $r\in[0,\ \infty)$, and we conclude $E_a\le E_b$. In the same manner we prove the following theorem:

\medskip

\noindent{\bf Theorem 5:} ~~{\it The potential $V$ belongs to one of the classes ${\it (1)}$--${\it (3)}$ and has $\psi_l r^{-sk_d}$--weighted area, $S=sV$, $sk_d<0$, and 
\begin{equation}\label{th9}
\mu(r)=\int_0^r(V_b(t)-V_a(t))|\psi_{l}(t)|t^{-sk_d}dt, \quad r\in [0,\ \infty).
\end{equation}
Then if $\mu\ge 0$, the eigenvalies are ordered, i.\,e. $E_a\le E_b$, where $\psi_l=\psi_{1i}$ if $s=1$ and $\psi_l=\psi_{2i}$ if $s=-1$, $i=a$ {\bf or} $b$.} 

\medskip

\noindent{\bf Proof:} We prove the theorem for the pseudo--spin symmetric case and assume that $\psi_2$ lies above the $r$-axis i.\,e. $s=-1$ and $\psi_2\ge 0$; for the other case the proof is similar. We integrate the right side of (\ref{expr7}) to obtain
\begin{equation*}
(E_b - E_a)\int_0^\infty (\psi_{1a}\psi_{1b} + \psi_{2a}\psi_{2b})dr=
-\int_0^\infty\mu\left(\psi_{2a}r^{-k_d}\right)'dr,       
\end{equation*}
where $\mu$ is defined by (\ref{th9}) for $i=b$. Since $\psi_{2a}\ge 0$, then $\psi_{2a} r^{k_d}\ge 0$ and, according to Lemma 2, $(\psi_{2a} r^{-k_d})'\le 0$. Thus the product $\mu\left(\psi_{2a}r^{-k_d}\right)'$ is nonpositive and the above expression leads to $E_a\le E_b$. 

\hfill $\Box$

\medskip

\noindent{\bf Corollary 5:} ~~{\it Let the comparison potentials belong to one of the classes ${\it (1)}$--${\it (3)}$. If the potentials cross over once, say at $r_1$, $V_a\le V_b$ for $r\in [0,\ r_1]$, and
\begin{equation*}
\mu(\infty)=\int_0^r (V_b-V_a)|\psi_{l}|r^{-sk_d}dr,
\end{equation*}
or if the potentials cross over twice, say at $r_1$ and $r_2$, $r_1<r_2$, $V_a\le V_b$ for $r\in [0,\ r_1]$, and
\begin{equation*}
\mu(r_2)=\int_0^{r_2} (V_b-V_a)|\psi_{l}|r^{-sk_d}dr.
\end{equation*}
Then if $\mu(\infty)\ge 0$ or $\mu(r_2)\ge 0$ it follows that $\mu(r)\ge 0$ and the eigenvalies are ordered, i.\,e. $E_a\le E_b$, where $\psi_l=\psi_{1i}$ if $s=1$ and $\psi_l=\psi_{2i}$ if $s=-1$, $i=a$ {\bf or} $b$.} 

\medskip

As before we can generalize Corollary 5 to allow $n$ intersections, i.e. if $V_a\le V_b$ on $r\in[0,\ r_1]$ and sequence of absolute areas $\int_{r_i}^{r_{i+1}}|(V_b-V_a)\psi_{l}r^{-sk_d}|dr$ is nonincreasing (if $n$ is odd then we assume $\int_{r_{n-1}}^{r_n}|(V_b-V_a)\psi_{l}r^{-sk_d}|dr\ge\int_{r_n}^\infty|(V_b-V_a)
\psi_{l}r^{-sk_d}|dr$), then integral $\int_0^r (V_b(t)-V_a(t))\psi_{l}(t)t^{-sk_d}dt\ge 0$ for $r\in[0,\ \infty)$, so $E_a\le E_b$.

%%%%%%%%%%%%%%%%%%%%%%%%%%%%%%%%%%%%%%%%%%%%%%%%%%%%%%%%%%%%%%%%%%%%%%%%%%%%%%%%%%%%%%%%
%%%%%%%%%%%%%%%%%%%%%%%%%%%%%%%%%%%%%%%%%%%%%%%%%%%%%%%%%%%%%%%%%%%%%%%%%%%%%%%%%%%%%%%%
\subsection{An example}
%%%%%%%%%%%%%%%%%%%%%%%%%%%%%%%%%%%%%%%%%%%%%%%%%%%%%%%%%%%%%%%%%%%%%%%%%%%%%%%%%%%%%%%%
%%%%%%%%%%%%%%%%%%%%%%%%%%%%%%%%%%%%%%%%%%%%%%%%%%%%%%%%%%%%%%%%%%%%%%%%%%%%%%%%%%%%%%%%
Here we will demonstrate the first part of Corollary 5, i.\,e. the case of one intersection. For the comparison potentials we choose the Yukawa potential \cite{Yuk} $V_a$ and the Coulomb potential $V_b$, which satisfy ${\it (1)}$ for $s=1$:  
\begin{equation*}
V_a=-\frac{\alpha}{re^{ar}} \qquad \text{and} \qquad 
V_b=-\frac{\beta}{r}.
\end{equation*}
The solutions of the Dirac Coulomb problem are well known; in particular, article \cite{HY} provides us with the eigenvalue equation and the ground state wave function for $d=2$, $j=1/2$, $\tau=-1$, and $m=1$ in the spin--symmetric case:
\begin{equation*}
E_b^2-1=-\left(2\beta(E_b+1)\right)^2 \quad \text{and} \quad
\psi_{b}=\sqrt{r}e^{-r\sqrt{1-E_b^2}}.
\end{equation*} 
Fixing $\alpha=0.2$ and $a=0.1$ and varying $\beta=0.172$, the above potentials intersect at exactly one point (Figure 3, left graph) so that $V_a\le V_b$ before the intersection point.  A direct numerical calculation shows
\begin{equation*}\label{concl5}
\mu(\infty)=\int_0^\infty (V_b-V_a)\psi_{b}\sqrt{r}dr=0.00006
\end{equation*}
Hence, according to Corollary 5, $E_a\le E_b$, which we have verified by an accurate calculation:  $E_a=0.75632\le E_b=0.78837$.
\begin{figure}[ht]
\begin{center}$
\begin{array}{cc}
\includegraphics[width=2.5in, angle=-90]{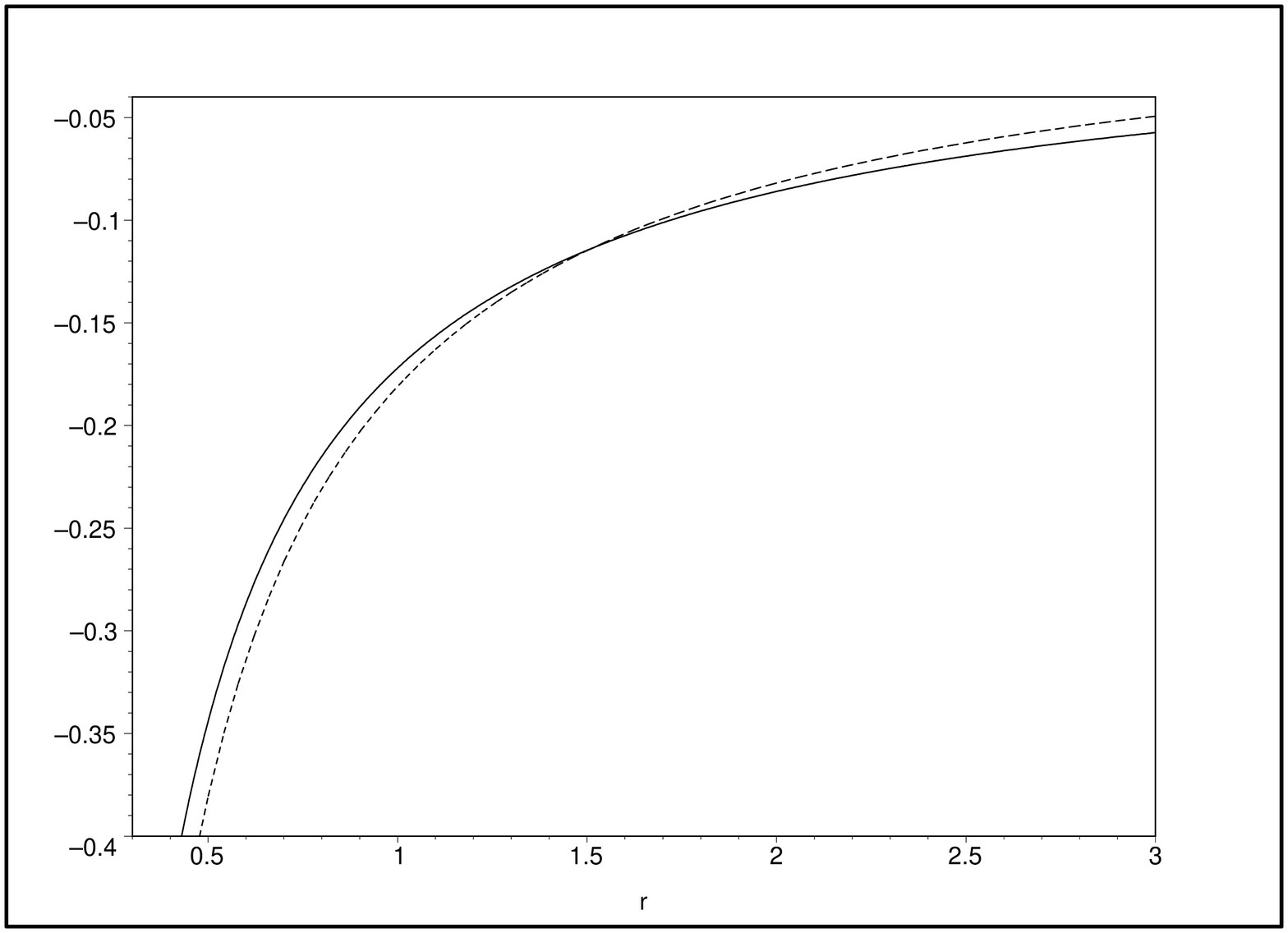} & 
\includegraphics[width=2.5in, angle=-90]{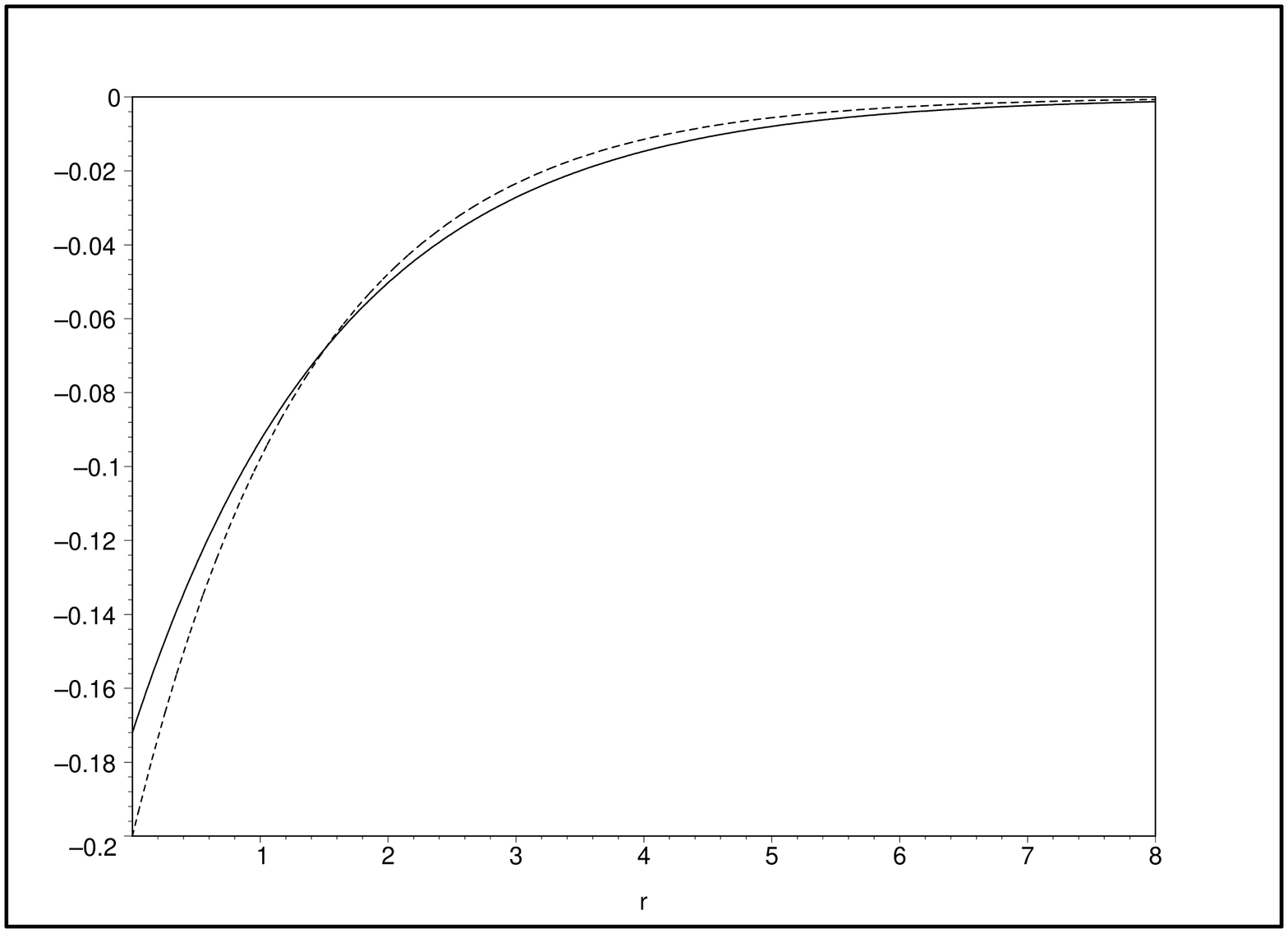}
\end{array}$
\end{center}
\caption{Left graph: The Yukawa potential $V_a$ (dotted line) and the Coulomb potential $V_b$ (full line). Right graph:  the functions $V_a\psi_{b}\sqrt{r}$ (dotted line) and $V_b\psi_{b}\sqrt{r}$ (full line).}
\end{figure}
\begin{figure}[ht]
\begin{center}$
\begin{array}{cc}
\includegraphics[width=2.5in, angle=-90]{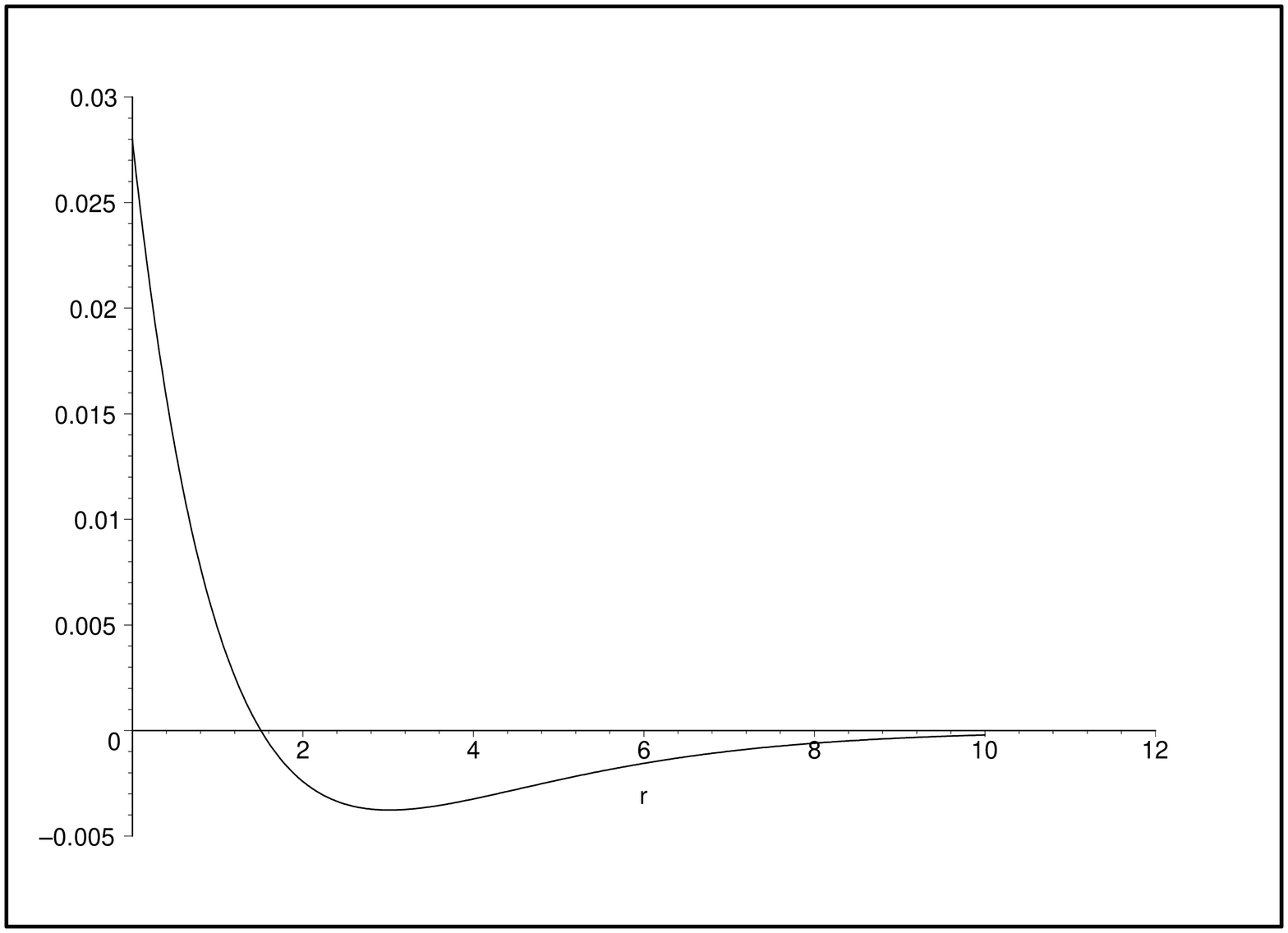} & 
\includegraphics[width=2.5in, angle=-90]{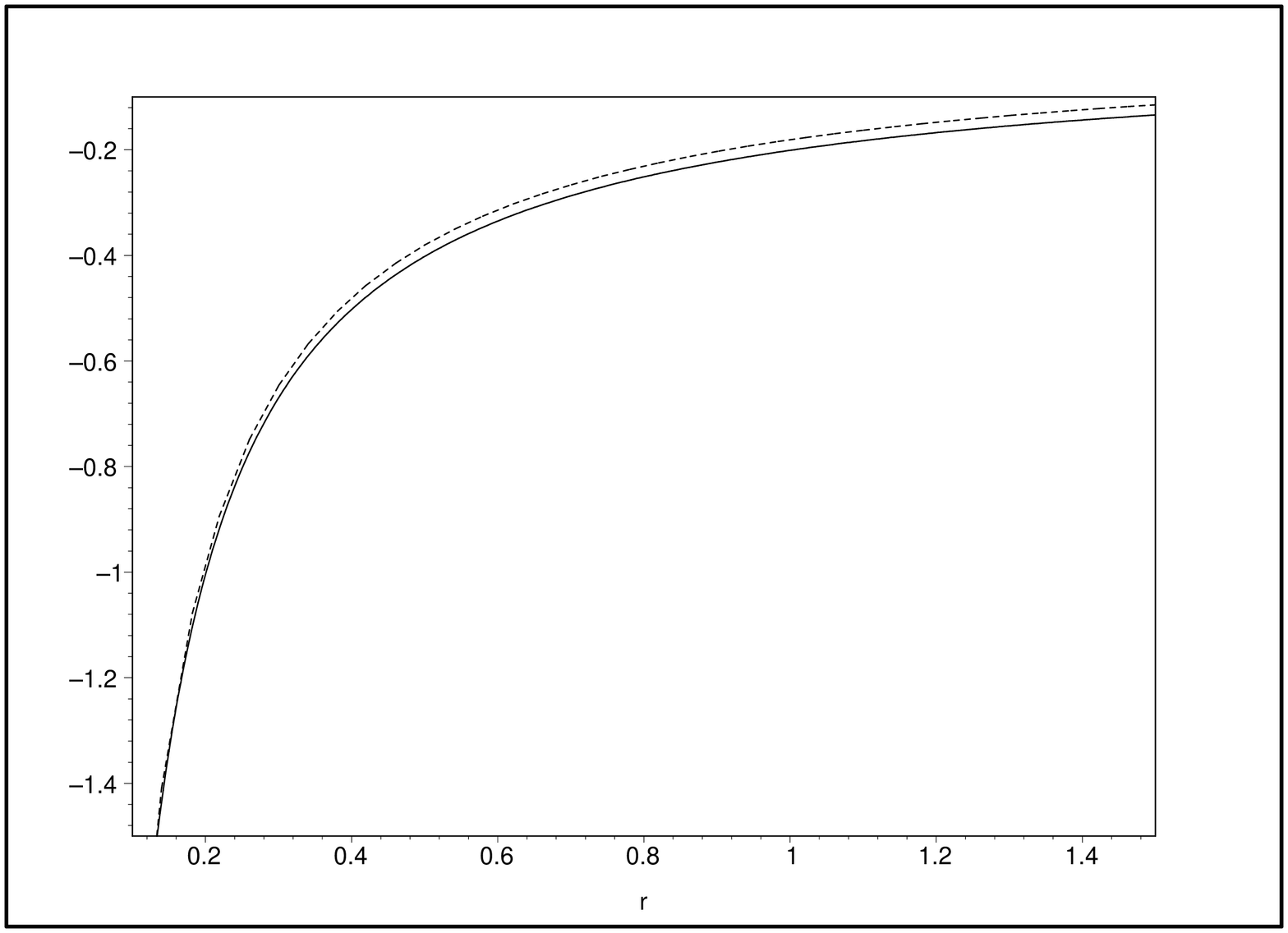}
\end{array}$
\end{center}
\caption{Left graph: The graph of the integrand $I=(V_b-V_a)\psi_{b}\sqrt{r}$. Right graph: The Coulomb potential $V_a$ (dotted line) and the Yukawa potential $V_b$ (full line).}
\end{figure}
We also can obtain a lower bound for $E_a$ using the usual comparison theorem, which follows from (\ref{expr7}). Keeping $\alpha=0.2$ and $a=0.1$ and choosing $\beta=0.201$ we have $V_a>V_b$ on $r\in[0,\ \infty)$ (see Figure 4, right graph). Therefore $E_a=0.75632>E_b=0.70010$.

%%%%%%%%%%%%%%%%%%%%%%%%%%%%%%%%%%%%%%%%%%%%%%%%%%%%%%%%%%%%%%%%%%%%%%%%%%%%%%%%%%%%%%%%
%%%%%%%%%%%%%%%%%%%%%%%%%%%%%%%%%%%%%%%%%%%%%%%%%%%%%%%%%%%%%%%%%%%%%%%%%%%%%%%%%%%%%%%%
\section{Conclusion}
%%%%%%%%%%%%%%%%%%%%%%%%%%%%%%%%%%%%%%%%%%%%%%%%%%%%%%%%%%%%%%%%%%%%%%%%%%%%%%%%%%%%%%%%
%%%%%%%%%%%%%%%%%%%%%%%%%%%%%%%%%%%%%%%%%%%%%%%%%%%%%%%%%%%%%%%%%%%%%%%%%%%%%%%%%%%%%%%%
The systems of Dirac coupled equations in one dimension (\ref{d1l})--(\ref{d1r}) and $d>1$ dimensions (\ref{Sl1})--(\ref{Sr1}) are studied here for the spin--symmetric and pseudo--spin--symmetric cases. The treatment of these two cases has been unified by the introduction of the parameter $s$ which takes the value $s=1$ if $S=V$ and $s=-1$ if $S=-V$, thus $S=sV$. By writing the above systems in a Schr\"{o}dinger--like form and analyzing their behaviour near the origin and at infinity, we able to consider three appropriate and interesting classes of potential, with corresponding general relations between energy $E$ and the mass of the particle $m$. The structure of the nodeless states were discussed, and certain monotone behaviours of the wave functions were established (Lemma 1 and Lemma 2). Using these results we have refined the comparison theorems for the Dirac equations in the $S=sV$ cases. In fact, the condition $V_a\le V_b$ which leads to $E_a\le E_b$ may now be replaced by $U_a\le U_b$, where in each case $U$ is a specific integral transform of $V$ that induces a weaker condition leading to the same spectral ordering $E_a\le E_b$.  For problems where it is found to be complicated to apply these theorems immediately, corresponding corollaries have been established for the cases of one, two, and $n$ intersections of the comparison--potential graphs.  The application of these theorems is illustrated by a variety of explicit examples in one and $d>1$ dimensions.  Since, exact analytical solutions  for spin--symmetric and pseudo--spin--symmetric  problems are plentiful in the literature, there is reason to expect that an approximation theory based on such comparison theorems might offer a useful tool for relativistic spectral estimation.

%%%%%%%%%%%%%%%%%%%%%%%%%%%%%%%%%%%%%%%%%%%%%%%%%%%%%%%%%%%%%%%%%%%%%%%%%%%%%%%%%%%%%%%%
%%%%%%%%%%%%%%%%%%%%%%%%%%%%%%%%%%%%%%%%%%%%%%%%%%%%%%%%%%%%%%%%%%%%%%%%%%%%%%%%%%%%%%%%
\section{Acknowledgements}
%%%%%%%%%%%%%%%%%%%%%%%%%%%%%%%%%%%%%%%%%%%%%%%%%%%%%%%%%%%%%%%%%%%%%%%%%%%%%%%%%%%%%%%%
%%%%%%%%%%%%%%%%%%%%%%%%%%%%%%%%%%%%%%%%%%%%%%%%%%%%%%%%%%%%%%%%%%%%%%%%%%%%%%%%%%%%%%%%
One of us (RLH) gratefully acknowledges partial financial support
of this research under Grant No.\ GP3438 from the Natural Sciences
and Engineering Research Council of Canada.\medskip

%%%%%%%%%%%%%%%%%%%%%%%%%%%%%%%%%%%%%%%%%%%%%%%%%%%%%%%%%%%%%%%%%%%%%%%%%%%%%%%%%%%%%%%%
%%%%%%%%%%%%%%%%%%%%%%%%%%%%%%%%%%%%%%%%%%%%%%%%%%%%%%%%%%%%%%%%%%%%%%%%%%%%%%%%%%%%%%%%
\section*{References}
%%%%%%%%%%%%%%%%%%%%%%%%%%%%%%%%%%%%%%%%%%%%%%%%%%%%%%%%%%%%%%%%%%%%%%%%%%%%%%%%%%%%%%%%
%%%%%%%%%%%%%%%%%%%%%%%%%%%%%%%%%%%%%%%%%%%%%%%%%%%%%%%%%%%%%%%%%%%%%%%%%%%%%%%%%%%%%%%%

\end{document}